\documentclass[aps,pra,twocolumn,floatfix, superscriptaddress, 10pt]{revtex4-2}
\usepackage[colorlinks=true, letterpaper=ture, pdfstartview=FitV, linkcolor=blue, citecolor=blue, urlcolor=blue]{hyperref}
\usepackage{amssymb, amsmath,color,mciteplus,graphicx,subfigure}
\usepackage{calc,epsfig,epstopdf,mciteplus,bm,mathrsfs}
\usepackage{times}
\usepackage{multirow}
\usepackage{lipsum}
\usepackage[ruled,vlined,linesnumbered]{algorithm2e}
\usepackage{dcolumn}
\usepackage{setspace}
\usepackage{xcolor}
\usepackage{bm}
\usepackage{ulem}

\usepackage{graphicx}

\begin{document}
    \title{Determining non-Hermitian parent Hamiltonian from a single  eigenstate}
	
    \author{Xu-Dan Xie}
    \affiliation{Key Laboratory of Atomic and Subatomic Structure and Quantum Control (Ministry of Education),\\  Guangdong Basic Research Center of Excellence for Structure and Fundamental Interactions of Matter,\\ and School of Physics, South China Normal University, Guangzhou 510006, China}
	
    \author{Zheng-Yuan Xue}  \email{zyxue@scnu.edu.cn}
	
    \author{Dan-Bo Zhang} \email{dbzhang@m.scnu.edu.cn}
    \affiliation{Key Laboratory of Atomic and Subatomic Structure and Quantum Control (Ministry of Education),\\  Guangdong Basic Research Center of Excellence for Structure and Fundamental Interactions of Matter,\\ and  School of Physics, South China Normal University, Guangzhou 510006, China}
    \affiliation{Guangdong Provincial Key Laboratory of Quantum Engineering and Quantum Materials,\\  Guangdong-Hong Kong Joint Laboratory of Quantum Matter, and Frontier Research Institute for Physics,\\  South China Normal University, Guangzhou 510006, China}
	
    \date{\today}

    \begin{abstract}
A quantum state for being an eigenstate of some local Hamiltonian should be constraint by zero energy variance and consequently, the constraint is rather strong that a single eigenstate may uniquely determine the Hamiltonian. For non-Hermitian systems, it is natural to expect that determining the Hamiltonian requires a pair of both left and right eigenstates. Here, we observe that it can be sufficient to determine a non-Hermitian Hamiltonian from a single right or left eigenstate. Our approach is based on the quantum covariance matrix, where the solution of Hamiltonian corresponds to the complex null vector. Our scheme favours non-Hermitian Hamiltonian learning on experimental quantum systems, as only the right eigenstates there can be accessed. Furthermore,  we use numerical simulations to examine the effects of measurement errors  and show the stability of our scheme. 
    
    \end{abstract}

    \maketitle
    \definecolor{RED}{RGB}{255,0,0}

    \section{Introduction} \label{sec:level1}    
Determining a physically parent Hamiltonian from a given quantum state, as a quantum inverse problem, has attracted attention recently~\cite{rattacaso2022optimal, rattacaso2024parent,shen2023construction,turkeshi2019entanglement,giudici2022locality,beau2021parent}. Addressing the quantum inverse problem offers insights into strategies for manipulating and preparing quantum states, as controlling quantum states requires the design of appropriate Hamiltonian of quantum systems~\cite{koch2022quantum,van2016optimal, georgescu2014quantum,franceschetti1999inverse,valenti2019hamiltonian,gentile2021learning,menke2021automated}. Furthermore, identification of parent Hamiltonians (PHs) is related to Hamiltonian learning~\cite{wiebe2014hamiltonian,wang2017experimental,granade2012robust} and holds significant utility in the verification of quantum devices~\cite{gheorghiu2019verification,eisert2020quantum,vsupic2020self,carrasco2021theoretical,kliesch2021theory}. Various methods have been developed to address the reconstruction of the parent Hamiltonian, allowing the recovery of PHs from single eigenstates \cite{qi2019determining,chertkov2018computational,greiter2018method,bairey2019learning,hou2020determining,cao2020supervised,chen2012ground,fernandez2015frustration,turkeshi2020parent} or time-dependent states \cite{zubida2021optimal,rattacaso2021optimal,hangleiter2021robustly,zhao2021characterizing,che2021learning,stilck2024efficient,rattacaso2023high}.
Currently, learning Hamiltonian from quantum states are often restricted to Hermitian quantum systems. However, a Hamiltonian description of quantum systems can also be non-Hermitian, which effectively encapsulates the interaction of the system with the environment. In fact, non-Hermitian physics is ubiquitous in nature and has been intensively studied \cite{ashida2020non}, including  open quantum systems \cite{dalibard1992wave,carmichael1993quantum,rotter2009non,daley2014quantum}, wave propagation processes under gain and loss \cite{feng2017non,el2018non,miri2019exceptional,ozawa2019topological}, non-unitary quantum field theories \cite{ fisher1978yang,alcaraz1987surface,korff2007pt} and so on. 
            
However, the inverse problem of leaning non-Hermitian Hamiltonian from the eigenstates has been hardly explored, with a few exceptions \cite{shen2023construction,tang2023non}. Recently, a method based on matrix product states was proposed to reconstruct the parent Hamiltonian from a pair of its ground states~\cite{shen2023construction}. Ref.~\cite{tang2023non} describes a numerical scheme to obtain the non-Hermitian local parent Hamiltonians  through generalized quantum covariance matrix~(QCM) method. These proposals require a pair of biorthogonal left- and right-eigenstates for reconstructing the non-Hermitian parent Hamiltonian. This is infeasible for Hamiltonian learning when only the right eigenstate can be accessed on real quantum systems. Then, a question naturally arises: can we reconstruct a non-Hermitian Hamiltonian solely from a single right/left eigenstate? At first glance, the answer should be negative since one may rebuild a Hermitian Hamiltonian from a single eigenstate. However, left or right eigenstates of local non-Hermitian systems can have different structures compared to eigenstates of local Hermitian systems \cite{martinez2018topological,cipolloni2023entanglement}. Consequently, when restricted to local Hamiltonians, an eigenstate may only serve as a right eigenstate for a non-Hermitian system due to its specific structure. 
        
In this paper, we present evidence that some non-Hermitian local Hamiltonian can be determined solely from a right or left eigenstate. Our approach for learning Hamiltonian from an eigenstate is based on the quantum covariance matrix, where the solution of Hamiltonian corresponds to the complex null vector of QCM. To demonstrate the effectiveness of our method, we select a non-Hermitian Ising spin chain and a random Hamiltonian as examples.  Numerical results closely match the theoretical values, indicating that it is possible to recover the local non-Hermitian parent Hamiltonian from a single right ( or left) eigenstate. Compared to the method in Ref.~\cite{tang2023non}, our covariance matrix has a simpler form. Additionally, our approach is easier to implement experimentally since it only requires information from the right eigenstate.  Our work uncovers the intrinsic relations between eigenstates and non-Hermitian Hamiltonians and can be experimentally relevant for identifying non-Hermitian quantum systems.

\section{ Quantum covariance matrix for determining Hamiltonian } \label{sec:qcm}
    
We first review the quantum covariance matrix method for determining the Hermitian Hamiltonian and its generalization for non-Hermitian Hamiltonian where both the left and right eigenstates should be involved. Then, we propose a quantum covariance matrix methof for determining a non-Hermitian Hamiltonian by using only the right or left eigenstate.

\subsection{Determining Hamiltonian from a pair of eigenstates }  
Given a Hamiltonian in quantum many-body physics, we can determine its eigenstates. We turn to its quantum inverse problem: given a quantum state $|v \rangle$, how can we determine its parent Hamiltonian, which has  $|v \rangle$ as an eigenstate? At first glance, this seems impossible because many Hamiltonian operators can share the same eigenstate. Without any constraint, the solution space for the inverse problem is increasing exponentially. However, a physically meaningful Hamiltonian is subject to some fundamental constraints. By imposing sufficient constraints, we can uniquely determine the parent Hamiltonian from the knowledge of a single eigenstate~\cite{qi2019determining}. 
    
One remarkable constraint is the locality of the Hamiltonian. When we confine the solution space to a local Hamiltonian, the parent Hamiltonian could be uniquely reconstructed \cite{qi2019determining}. For a lattice system with $n$ sites, the local Hamiltonian  can be written as a linear combination of local operators $\hat{H}(\bm{\omega})=\sum_i \omega_i \hat{O}_i$, where $\{\hat{O}_i \}$ is a set of range-$k$ local Hermitian operator basis with $k\ll n$.  In the Hermitian case, the coefficient $\bm{\omega}=[\omega_0,\omega_1,\dots,\omega_n]^T$ is a set of real numbers. With the constraint of locality, the quantum inverse problem is transformed into finding the corresponding coefficient $\bm{\omega}$ to satisfy the eigen equation $\hat{H}(\bm{\omega}) |v\rangle= \lambda |v\rangle $. 
    
A Hermitian Hamiltonian has the eigenstate $|v \rangle$ if and only if the energy variance is zero. The zero-energy variance imposes a constraint for the parent Hamiltonian. It is convenient to introduce the quantum covariance matrix (QCM), which is defined as \cite{qi2019determining}
             \begin{eqnarray}\label{Eq:M^v}
                M_{ij}^{v}=\frac{1}{2}\langle \hat{O}_i \hat{O}_j+\hat{O}_j \hat{O}_i \rangle_v -\langle \hat{O}_i \rangle_v \langle \hat{O}_j \rangle_v,
             \end{eqnarray}
where $\langle \cdot \rangle_v$  denotes $\langle v| \cdot |v \rangle$. The QCM $M^v$ can be used to compute the energy variance of $\hat{H}(\bm{\omega})$ in the state $|v \rangle$   
             \begin{eqnarray}\label{Eq:var_her}
                 \sigma^2 =\langle \hat{H}^2(\bm{\omega}) \rangle_v -\langle \hat{H}(\bm{\omega}) \rangle_v^2 
                  =\bm{\omega}^{T} \bm{M}^v \bm{\omega} \geq 0.
             \end{eqnarray}
According to Eq. (\ref{Eq:var_her}), zero-variance implies $\bm{\omega}^{T} \bm{M}^v \bm{\omega} = 0$, which can be deduced to $\bm{M}^v \bm{\omega} = 0$, as $\bm{M}^v$ is a positive-semidefinite matrix. It means that the coefficient $\bm{\omega}$ is corresponding to the eigenvector of  $\bm{M}^v$ with eigenvalue 0. Therefore, the null space of $\bm{M}^v$ is the solution space of the quantum inverse problem. Then we can get the solution by diagonalizing $\bm{M}^v$. Moreover, the solution of $\bm{\omega}$ should be real, ensuring that the corresponding Hamiltonian is Hermitian~\cite{qi2019determining}.
    
In Ref.~\cite{tang2023non}, the QCM method has been generalized to the non-Hermitian Hamiltonian, that is, $\hat{H}\neq \hat{H}^{\dag}$. In non-Hermitian case, $\hat{H}$ and $\hat{H}^{\dag}$ have distinct sets of eigenstates,       
    	\begin{eqnarray}
    		\hat{H}\left | R_i  \right \rangle=E_{i} \left | R_i  \right \rangle, \quad
    		\hat{H}^\dag \left | L_i  \right \rangle=E^{*}_{i} \left | L_i  \right \rangle, 
    		\label{eq:eigen}
    	\end{eqnarray}
where $| R_n \rangle $ is called the right eigenstate and $| L_n  \rangle $ is called the left eigenstate. In the biorthogonal basis, the non-Hermitian Hamiltonian can be diagonalized as follows        
    	\begin{eqnarray}
\hat{H}=\sum_i E_i  | R_i   \rangle \langle   L_i  |.
    		\label{eq:diag}
    	\end{eqnarray}
The local non-Hermitian Hamiltonian can also be written as $\hat{H}(\bm{\omega})=\sum_i \omega_i \hat{O}_i$. Note that the coefficients $\bm{\omega}=[\omega_0,\omega_1,\dots,\omega_n]^T$ is a set of complex numbers, which is different from the Hermitian case.  To reconstruct the local non-Hermitian parent Hamiltonian,  a general quantum covariance matrix is proposed~\cite{tang2023non}, where both the left and right eigenstates are used,    
           \begin{eqnarray}
             C_{ij}^{LR}=&& \frac{\langle R|\hat{O}_j^\dagger(1-|L\rangle\langle R|)(1-|R\rangle\langle L|)\hat{O}_i|R\rangle}{2\langle R|R\rangle} \\ \notag
              &&+\frac{\langle L|\hat{O}_i(1-|R\rangle\langle L|)(1-|L\rangle\langle R|)\hat{O}_j^\dagger|L\rangle}{2\langle L|L\rangle}
           \end{eqnarray}
The general QCM is a Hermitian and positive-semidefinite matrix. Similar to the Hermitian case, the null space of the QCM $C^{LR}$ is exactly the solution space of the non-Hermitain quantum inverse problem.

\subsection{Determining Hamiltonian from a single eigenstate} 

The use of biorthogonal eigenstates is a common approach in identifying non-Hermitian systems, i.e., one can use the general QCM to recover the non-Hermitian parent Hamiltonian from a pair of  biorthogonal eigenstates $| R \rangle$ and $| L \rangle$. However, when one can only access the right eigenstate on quantum devices, the general QCM approach will fail to work. It is valuable to relax the condition of a single pair of biorthogonal eigenstates to a single right or left eigenstate. Is it possible to determine the non-Hermitian parent Hamiltonian solely based on a single left or right eigenstate? 
         
Let us begin by introducing the variance of non-Hermitian operators. In non-Hermitian systems, the definition formula of variance is slightly different from that in Hermitian systems. The variance of a non-Hermitian operator $\hat{A}$ in a quantum state $|\psi\rangle$ is defined as \cite{pati2015measuring,percival1998quantum}:
            \begin{eqnarray} \label{Eq:non_var}
                \triangle A^2&&=\langle \psi |(\hat{A}^{\dagger}-\langle \hat{A}^{\dagger}\rangle)(\hat{A}-\langle \hat{A}\rangle)|\psi \rangle \\ \notag 
                &&=\langle \psi |\hat{A}^{\dagger} \hat{A}|\psi \rangle-\langle \psi |\hat{A}^{\dagger} |\psi \rangle \langle \psi | \hat{A}|\psi \rangle\\ \notag
                &&=\langle f |f \rangle \geq 0,
            \end{eqnarray}
where $\langle \hat{A}\rangle=\langle \psi |\hat{A}|\psi \rangle$ , $\langle \hat{A}^{\dagger}\rangle=\langle \psi |\hat{A}^{\dagger}|\psi \rangle$ and $|f \rangle=(\hat{A}-\langle \hat{A}\rangle)|\psi \rangle$. Obviously, if and only if the state $|\psi\rangle$ is an right eigenstate of $\hat{A}$, the variance has a minimum value of 0. By utilizing the property of zero variance, we can calculate the eigenstates of the Hamiltonian operators ~\cite{zhang2022variational,chen2023variational,xie2024variational}. Similarly, we can determine the parent Hamiltonian from their eigenstates.
    
According to the Eq. (\ref{Eq:non_var}), the energy variance of the local non-Hermitian Hamiltonian $\hat{H}$ in the quantum state $|R\rangle$ is written as        
            \begin{eqnarray}\label{Eq:energy_var}
                \triangle H^2=\langle H^\dagger H\rangle_R-\langle H^\dagger \rangle_R \langle H\rangle_R
            \end{eqnarray}
Now we define QCM $M^R$ as 
            \begin{eqnarray}\label{Eq:M_R}
                M^R_{ij}= \langle \hat{O}_i \hat{O}_j \rangle_R- \langle \hat{O}_i \rangle_R \langle \hat{O}_j \rangle_R.
            \end{eqnarray}
The elements of QCM are composed of the expected values of the Hermitian operator in the state $|R\rangle$, which can be obtained by measuring in experiments.        Then the energy variance of $\hat{H}(\bm{\omega})=\sum_i \omega_i \hat{O}_i$ can be expressed as        
           \begin{eqnarray} \label{Eq:var_non}
                \triangle H^2&&= \sum_{ij} \omega_i^* \omega_j [\langle \hat{O}_i \hat{O}_j \rangle_R -\langle \hat{O}_i \rangle_R \langle \hat{O}_j \rangle_R] \\ \notag
                &&= \sum_{ij} \omega_i^* M^R_{i,j}\omega_j \\ \notag
                && =\bm{\omega}^\dagger \bm{ M}^R   \bm{\omega} \geq 0
            \end{eqnarray}
If the state $|R\rangle$ is the eigenstate of the Hamiltonian $\hat{H}(\bm{\omega})$, the energy variance should be zero. In other words, to find the parent Hamiltonian of $|R\rangle$, one must locate $\bm{\omega}$ that satisfies the condition $\bm{\omega}^\dagger \bm{M}^R \bm{\omega} = 0$. The Hamiltonian determination thus becomes a linear algebra problem.
    
According to Eq.~\eqref{Eq:var_non}, $M^R$ is a positive semi-definite matrix. Thus, $M^R$ can be decomposed as      
            \begin{eqnarray}
                M^R = \sum m_i |S_i\rangle \langle S_i|,
            \end{eqnarray}
where $|S_i\rangle$ are eigenvectors of $M^R$ with eigenvalues and the spectrum $\{m_i\}$ satisfies          
             \begin{eqnarray}
                 0 \leq m_0 \leq m_1, \ldots, 
                 \end{eqnarray}
which can be called the correlation spectrum.
In the basis of $\{|S_i\rangle\}$, $\bm{\omega}$ can be written as $\bm{\omega} = \sum a_i |S_i\rangle$. 
Consequently,            
            \begin{eqnarray}
                \bm{\omega}^\dagger \bm{M}^R \bm{\omega} &=& \sum_{ijk} a_j^* \langle S_j | m_k |S_k \rangle \langle S_k | a_i |S_i \rangle \\ \notag
                &=& \sum_{ijk} a_j^* a_i m_k \langle S_j | S_k \rangle \langle S_k | S_i \rangle \\ \notag
                &=& \sum_k |a_k|^2 m_k.
            \end{eqnarray}        
To make the variance zero, either $m_k = 0$ when $a_k \neq 0$, or $a_k = 0$ when $m_k \neq 0$. In other words, $\bm{\omega}$ must belong to the null space of $M^R$. 
    
       
Therefore, the eigenvector of the QCM $M^R$ with the zero eigenvalue is exactly the coefficient set $\bm{\omega}$ desired to reconstruct the parent Hamiltonian. If $M^R$ has no eigenvector with zero eigenvalue, there is no eigenstate of any local Hamiltonian. If $M^R$ has only one zero eigenvalue, we can uniquely determine the local non-Hermitian Hamiltonian. 
(Here we consider $\hat{H}(\bm{\omega})$ and $\alpha \hat{H}(\bm{\omega})+\beta$ as the same Hamiltonian, as the addition of a constant term $\beta$ and scaling by a factor $\alpha$ does not change the underlying physics of the system.) If $M^R$ has more than one zero eigenvalue, then the coefficients we want to solve are the linear combinations of these eigenvectors with zero eigenvalues. So we can obtain the solution by diagonalizing the QCM $M^R$. 
Remarkably, once the Hamiltonian is local, the dimension of QCM is only increasing polynomially with the system size. This requires only a polynomial number of measurements for the QCM, and solving the null vector of QCM can also be efficient.  
    
The QCM $M^R$ shares a formula similar to $M^v$ in Eq. (\eqref{Eq:M^v}), except that it uses a quantum average of $\hat{O}_i \hat{O}_j$ instead of a symmetric one $\hat{O}_i \hat{O}_j+\hat{O}_j \hat{O}_i$. One remarkable difference is that the null vector for $M^R$ can be complex, leading a solution of non-Hermitian Hamiltonian. 
        
In the same way, we can use the QCM method to recover the parent Hamiltonian from the left eigenstate $|L\rangle$. In this case, QCM $M^L$ is defined as
           \begin{eqnarray}\label{Eq:M_L}
                M^L_{ij}= \big(\langle \hat{O}_i \hat{O}_j \rangle_L- \langle \hat{O}_i \rangle_L \langle \hat{O}_j \rangle_L \big)^*.
           \end{eqnarray}
As in the case of the right eigenvector, $M^L$ is a positive semi-definite matrix, and its null space allows us to rebuild the non-Hermitian parent Hamiltonian.


\section{numerical results} \label{sec:numeral}
In this section, we test our QCM approach introduced in Eq.~\eqref{Eq:M_R} for determining the non-Hermitian local Hamiltonian with some physical states by numerical simulations. 

We chose $|R\rangle$ as a physical state, e.g. obtained as an eigenstate for some local Hamiltonian, rather than some random states. To demonstrate the effectiveness of our QCM method, we select a prototypical non-Hermitian Ising model~\cite{von1991critical},        
            \begin{eqnarray}
                H_{\lambda,\kappa}=\sum_j^N (\lambda \sigma_j^x\sigma_{j+1}^x+ g \sigma_j^z+\text{i} \kappa \sigma_j^x),
            \end{eqnarray}
with an imaginary longitudinal field of strength $\kappa$, where $\lambda,g,\kappa $ are real numbers.  Since the Hamiltonian only has nearest-neighbor interactions, the operator set can be chosen among the rank-2 local operators,        
             \begin{eqnarray}
                 \{\hat{O}\}=\{ \sigma^q_i,\sigma^q_i\sigma^p_{i+1}\ | i=1,2\dots N; p,q=x,y,z \},
             \end{eqnarray}
 which contains approximately $12N$ operators. Assume the optimal parent Hamiltonian is a combination of the rank-2  local operators, $\hat{H}(\bm{\tilde{\omega}})=\sum_i \tilde{\omega}_i \hat{O}_i$. Our aim is to reconstruct the parent Hamiltonian from a single right (left) eigenstate, implying the identification of the optimal coefficient set $\bm{\tilde{\omega}}$ that exactly corresponds to the null eigenvectors of the QCM.

        \begin{figure}[tp]
            \centering
            \includegraphics[width=\linewidth]{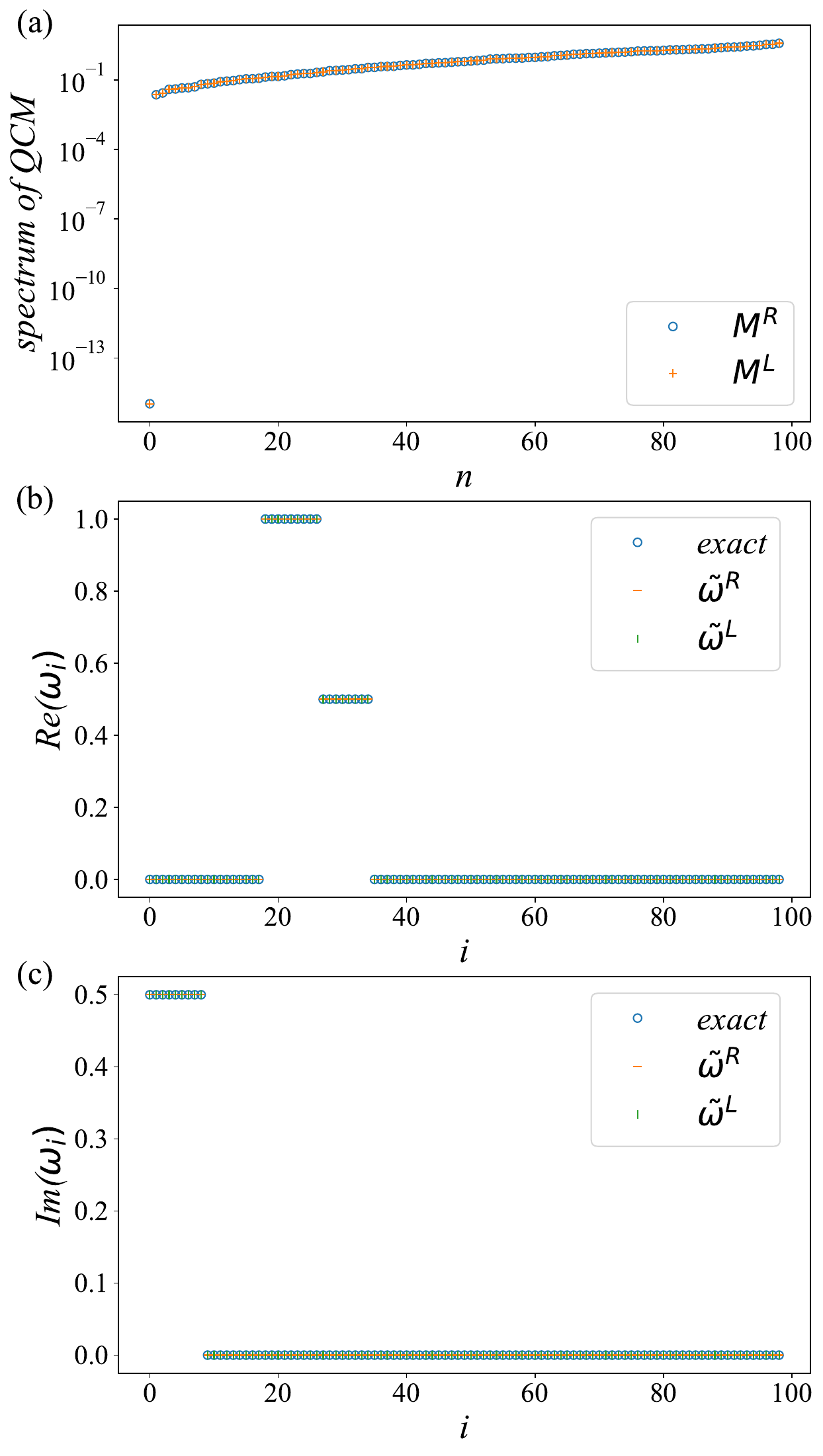}
            \caption{ Numerical result of the QCM method for $H_{\lambda,\kappa}$ in the case $\lambda=0.5,g=1,\kappa=0.5 $. (a) The eigenvalue spectrum of the generalized QCM; (b)  the real  part of the null eigenvector compared to the original Hamiltonian parameters with $\lambda=0.5,g=1,$and others set to 0; (c) the imaginary part of the null eigenvector compared to the original  Hamiltonian parameters,with $\kappa=0.5$, and others set to 0. } 
                \label{fig:QCM}
        \end{figure}

To begin with, we obtain all the biorthogonal eigenstates of the Hamiltonian $H_{\lambda,\kappa}$ through diagonalization. Then we randomly select a single pair of biorthogonal eigenstates  $|R\rangle$ and   $|L\rangle$, and construct the QCM $M^R$ and $M^L$ according to Eq.~\eqref{Eq:M_R} and Eq.~\eqref{Eq:M_L}, respectively. After spectral decomposition, we can get the spectrum and the null eigenvector of the QCM. The null eigenvectors of the QCM $M^R$ and $M^L$ are respectively denoted as $\bm{\tilde{\omega}}^R$ and $\bm{\tilde{\omega}}^L$.

\begin{figure}[tp]
                \centering
                \includegraphics[width= \linewidth]{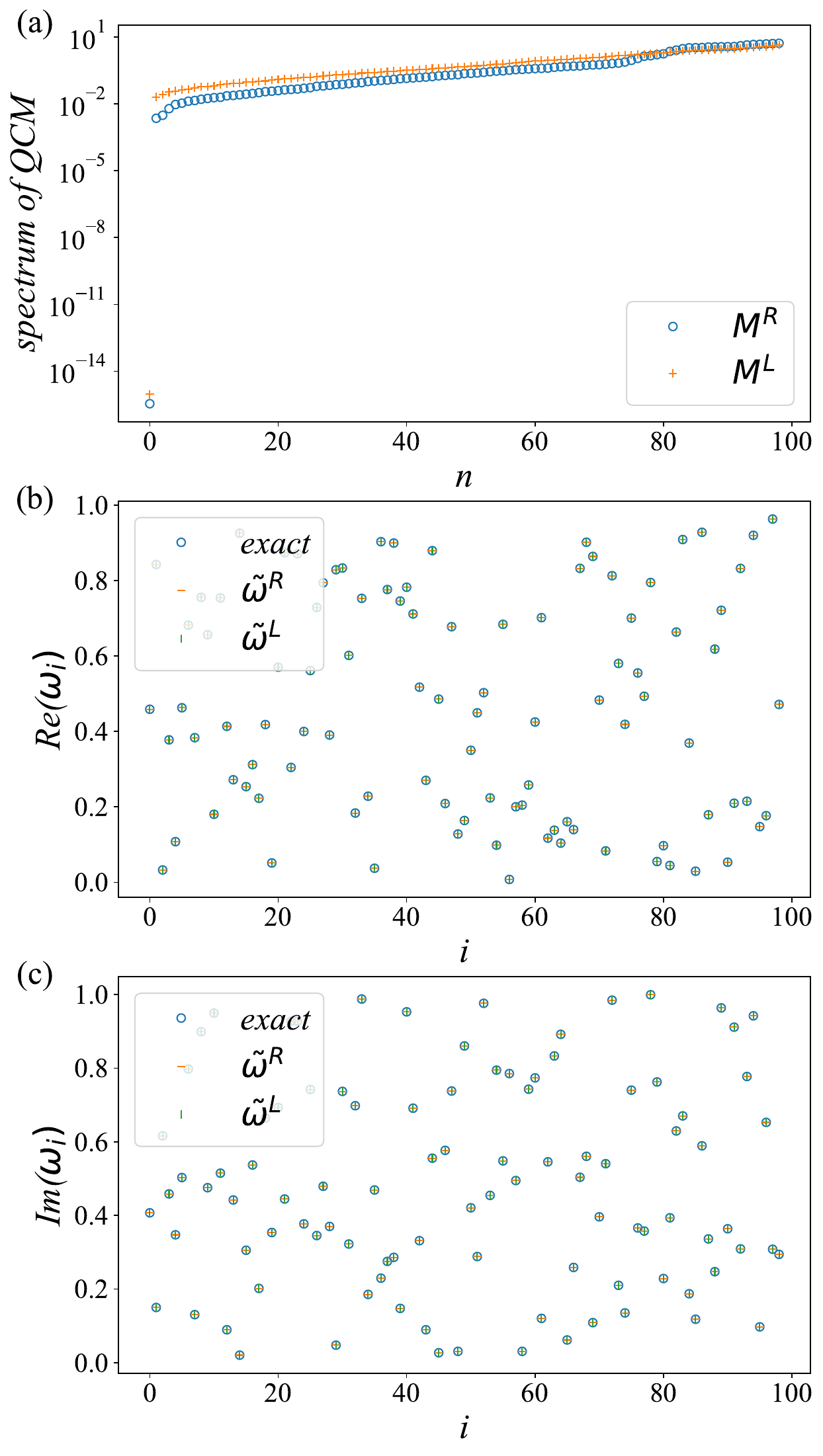}
                \caption{ Numerical result of the QCM method for the local non-hermitian Hamiltonian with random parameters. (a) the eigenvalue spectrum of the generalized QCM; (b)  the real  part of the null eigenvector compared with the coefficients for original Hamiltonian parameters; (c) the imaginary part of the null eigenvector compared with the coefficients for original Hamiltonian parameters. } 
                \label{fig:QCM_random}
            \end{figure}
    
Fig. \ref{fig:QCM} shows that the eigenvalue spectrum of the QCM obtained $M^R$ and $M^L$ both have a zero eigenvalue ($< 10^{-13}$). This indicates that from a single right (left) eigenvector, we can uniquely determine the local non-Hermitian parent Hamiltonian. In Fig. \ref{fig:QCM}, we display the real and imaginary components of $\bm{\tilde{\omega}}^R$ and $\bm{\tilde{\omega}}^L$ individually, juxtaposed against the original Hamiltonian parameters. It is evident that they exhibit a high degree of consistency with the original parameters, indicating the successful recovery of the parent Hamiltonian from a single right (left) eigenstate. Another noteworthy observation from Fig. \ref{fig:QCM} (a) is that the QCM $M^R$ and $M^L$ derived from a pair of biorthogonal eigenstates  possess an identical eigenvalue spectrum. This is because the Hamiltonian $H_{\lambda,\kappa}$ is transposed invariant. For non-Hermitian Hamiltonians with this symmetry, the left and right eigenvectors are found to be conjugate to each other, indicating a duality relationship between them. Therefore, in this case, the information of the system obtained from the left eigenstate is the same as that obtained from the right eigenstate.  
    
We further demonstrate with a local non-hermitian Hamiltonian with random parameters, $\hat{H}=\sum_i\omega_i \hat{O}_i$, where $\omega_i$ us a random complex number with both its real and imaginary parts lying within the interval $[0,1]$. 
We randomly choose a pair of  biorthogonal eigenstates of this Hamiltonian as input and use QCM method to recover the parent Hamiltonian. Due to the absence of transpose symmetry in this randomly generated Hamiltonian, the eigenvalue spectra of the QCM derived from the left and right eigenstates are different, as shown in Fig. \ref{fig:QCM_random}(a). However, we can retrieve the original Hamiltonian with the left or right eigenvectors, as shown in Figs. \ref{fig:QCM_random}(b) and \ref{fig:QCM_random}(c). This observation implies that, for general local Hamiltonians, the QCM method works well.

To demonstrate the robustness of our method, we utilize the standard deviation to portray the error of the QCM method, 
            \begin{eqnarray}\label{eq:error}
                \sigma(\bm{\omega'},\bm{\omega})=\sqrt{\frac{1}{K}\sum_i^K |\omega'_i-\omega_i|^2}
            \end{eqnarray}
where $\bm{\omega'}$ is the original Hamiltonian parameter and the $\omega$ is Hamiltonian parameter obtained by the QCM method.

            \begin{figure}[t]
                \centering
                \includegraphics[width= \linewidth]{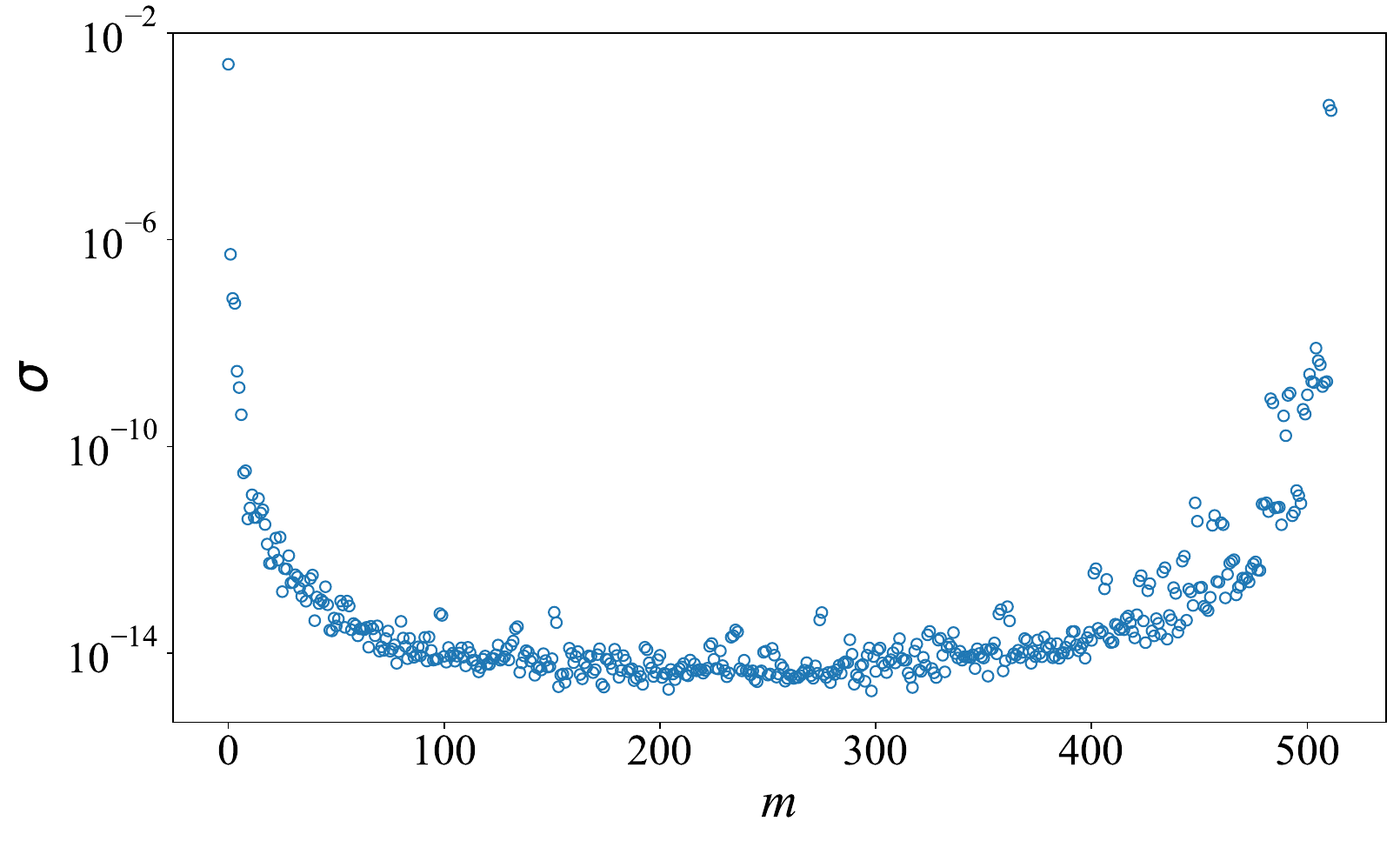}
                \caption{ The error of the QCM method for $H_{\lambda,\kappa}$ in the case $\lambda=0.5,\kappa=0.5 $. The graph shows the errors between $\bm{\tilde{\omega}}^R$ obtained by the QCM method and the original ones, in different right eigenstates of $H_{\lambda,\kappa}$. The horizontal axis represents the $m$th right eigenstate of $H_{\lambda,\kappa}$.} 
                \label{fig:error}
            \end{figure}

    \begin{figure} [t]
            \centering
            \includegraphics[width=1\linewidth]{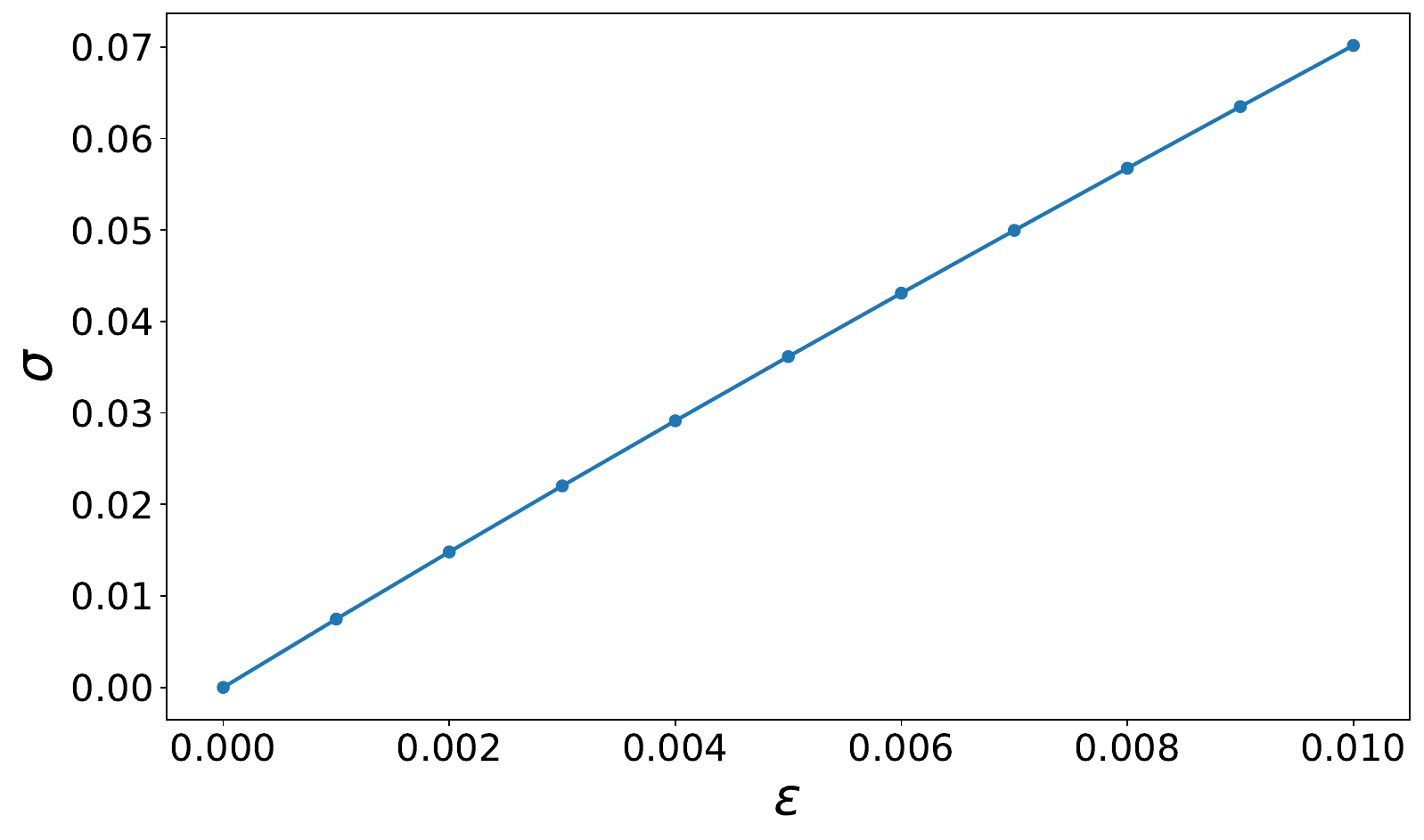}
             \caption{The error of the QCM numerically calculated in different measurement error. We select  512-th eigenstate of $H_{\lambda,\kappa}$ as trial state with $L=10$. The error matrix $\Delta M$ is generated by random numbers.} 
            \label{fig:measure_error} 
        \end{figure}

Fig. \ref{fig:error}  showcases the discrepancies between the original Hamiltonian and the reconstructed parent Hamiltonian, which are derived from various eigenstates of $H_{\lambda,\kappa}$. It is observed that the errors are always below $10^{-2}$ for all eigenstates of $H_{\lambda,\kappa}$. This suggests that the local non-Hermitian Hamiltonian can be sufficiently understood from a single right (left) eigenstate.  
An interesting observation is that the error is generally lower for eigenstates located in the middle of the spectrum (mid-spectrum) compared to those at the edges. This phenomenon aligns with the understanding that mid-spectrum eigenstates typically exhibit a larger entanglement entropy, following volume-law scaling, thus conforming to the Eigenstate Thermalization Hypothesis ~\cite{haque2022entanglement}. Eigenstates that adhere to the hypothesis encode the full Hamiltonian \cite{garrison2018does}, allowing for a more precise reconstruction of the parent Hamiltonian.

In practical applications,  when we aim to learn the parent Hamiltonian of a quantum system,it is necessary to perform measurements to construct the QCM,  which inherently introduces measurement errors. In this case, the covariance matrix is given by e
            \begin{eqnarray}
                M^R_1 =M^R+\varepsilon \Delta M 
            \end{eqnarray}
where $ \varepsilon $ is a character of deviation and $\Delta M$ represents the measurement errors.
    
For convenience, we denote the $i$-th eigenvector of $M^R$ with eigenvalue $\lambda_i$ as $\bm{\omega_i}$ which satisfies $\bm{\omega_i}^\dag\bm{\omega_j}=\delta_{ij}$ in  normalized conditions. The zero eigenvector of $M^R_1$ is denoted as $\bm{\omega'_0}$ (If $M^R_1$ has no zero eigenvector, we choose the eigenvector  with its smallest eigenvalue). Using standard non-degenerate perturbation theory in quantum mechanics, the relationship between $\bm{\omega'_0}$ and $\bm{\omega_0}$ is given by        
            \begin{eqnarray} \label{eq:omega_error}  
                \bm{\omega'_0} \propto \bm{\omega_0}+\varepsilon \sum_{i>0} \frac{\bm{\omega_i}^\dag \Delta M \bm{\omega_0}}{\lambda_i-\lambda_0} \bm{\omega_i}+O(\varepsilon^2).
            \end{eqnarray}             
We use one of the mid-spectrum eigenstate of $H_{\lambda,\kappa}$ with $L=10$ as trial state. Under different measurement errors, we computed the QCM. According to Eq. (\ref{eq:error}), we compute the error of QCM . Fig. \ref{fig:measure_error} illustrates the scaling of the error with respect to $\varepsilon$, which aligns with Eq. (\ref{eq:omega_error}). The deviation between the obtained parent Hamiltonian and the original Hamiltonian grows linearly with the measurement error when  $ \varepsilon \ll 1$.
     
\section{conclusion}\label{sec:conclusion}
    
In summary, we have introduced a QCM approach that enables the recovery of non-Hermitian local parent Hamiltonians from a single right eigenstate. By using the non-Hermitian Ising model and a random Hamiltonian as examples, we have demonstrated with numerical simulations  the effectiveness of our method. The results have shown that mid-spectrum eigenstates produce more accurate results than those at the edge of the spectrum. 
Our work not only implies a strong constraint on the right or left eigenstate and the non-Hermitian Hamiltonian, but also suggests a natural choice for learning as a non-Hermitian Hamiltonian from states on quantum devices.

\acknowledgments
    
This work was supported by the National Natural Science Foundation of China (Grant No.12375013 and  No.12275090), and the Guangdong Basic and Applied Basic Research Fund (Grant No.2023A1515011460). 

        \normalem	
\bibliography{reference}
	
\end{document}